\documentclass[prd,showpacs,preprint]{revtex4}
\usepackage{amsmath}
\usepackage{graphicx}
\usepackage{braket}

\begin{document}

\title{New constraints on R-parity violating couplings through the measurements of the $B^0_{s(d)}$-${\bar{B}^0_{s(d)}}$ and $K^0$-$\bar{K}^0$ mixing}
\author{Gao Xiangdong}
\email{gaoxiangdong@pku.edu.cn}
\author{Chong Sheng Li}
\email{csli@pku.edu.cn}
\author{Li Lin Yang}
\email{llyang@pku.edu.cn}
\affiliation{Department of Physics, Peking University, Beijing 100871, China}

\date{\today}

\begin{abstract}

We calculate contributions to $B_s-\bar{B}_s$ mixing through
tree-level sneutrino exchange in the framework of the minimal
supersymmetric standard model with R-parity violation, including the
next-to-leading-order QCD corrections. We compare our results with
the updated bounds on the $B_s-\bar{B}_s$ mass difference reported
by CDF collaborations, and present new constraints on the relevant
combinations of parameters of the minimal supersymmetric standard
model with R-parity violation. Our results show that upper bound on
the relevant combination of couplings of $B_s-\bar{B}_s$ mixing is
of the order $10^{-9}$. We also calculate the $B^0_d-\bar{B}^0_d$
and $K^0-\bar{K}^0$ mass differences, and show that the upper bounds
on the relevant combinations of couplings are two and four orders of
magnitude stronger than ones reported in the literatures,
respectively. We also discuss the case of complex couplings and show
that how the relevant combinations of couplings are constrained by
the updated experiment data of $B_s-\bar{B}_s$, $B_d-\bar{B}_d$
mixing and time-dependent CP asymmetry $S_{J/\psi K_s}$, and future
possible observations of $S_{J/\psi\phi}$ at LHCb, respectively.

\end{abstract}


\maketitle

\section{introduction}

Very recently, the {D\O} collaboration and the CDF collaboration at
the Fermilab Tevatron reported their updated measurements of the
mass difference between $B_s$ and $\bar{B}_s$ mesons ($\Delta m_s$).
The new bounds on the mass difference are~\cite{exp1,exp2}:
\begin{align}
  &\text{D\O:} \quad 17~\text{ps}^{-1} < \Delta m_s < 21~\text{ps}^{-1} \nonumber
  \\
  &\text{CDF:} \quad \Delta m_s = 17.77^{+0.10}_{-0.10} \pm 0.07~\text{ps}^{-1}
  \label{FBsmixing}
\end{align}
It was the first time that both the lower bound and the upper bound
for the $B_s-\bar{B}_s$ mixing are presented. Especially the CDF
result has reached an accuracy of about 1\%. The new results are
important for the precision test of the standard model (SM),
especially for the determination of the unitary triangle. Moreover,
if the SM predictions are consistent with the above results, these
data will put severe constraints on the flavor structure of the
possible new physics models beyond the SM.

In the literature, there have already been many discussions about
the implications of the new measurements. In
Ref.~\cite{buras2,Ligeti:2006pm,Ball:2006xx,Alakabha Datta,UTfit},
the authors carried out model-independent analysis of the
constraints on extensions of the SM. There are also model-dependent
calculations in some new physics models beyond the SM
\cite{FVMSSM,LHM,Zprime,Zprime2,GUT,Paradisi,Jonparry}. In this
paper, we further investigate the effects of the minimal
supersymmetric standard model with R-parity violation
(MSSM-RPV)~\cite{RPV} on the neutral meson mixing.

R-parity is a discrete symmetry defined by $R_p=(-1)^{3B+L+2S}$,
where B is the baryon number,L is the lepton number and S is the
spin of the particle. In a supersymmetric extension of the SM, all
the particles in the SM have $R_p=1$, while all the superpartners
have $R_p=-1$. R-parity conservation is imposed in the minimal
supersymmetric standard model (MSSM) to keep proton stable. However,
this requirement is not necessary for a fundamental theory, and one
can always introduce lepton number or baryon number violating terms
in the Lagrangian. For the neutral meson mixing, the relevant terms
in the Lagrangian are
\begin{equation}
  \mathcal{L}_{\text{RPV}} = \lambda'_{ijk} \tilde{\nu}_{iL}
  \bar{d}_{kR} d_{jL} + h.c. \label{largrangian}
\end{equation}
These interaction terms can induce $B_s-\bar{B}_s$ mixing through
tree level sneutrino exchange, and thus, probably, the relevant
combination of couplings $\sum\limits_i \lambda'_{i32}
\lambda'^*_{i23}$ will be severely constrained by the recent
measurements\cite{exp1,exp2}. Similarly, the terms shown in Eq.
(\ref{largrangian}) also can induce the $B^0_d-\bar{B}^0_d$ and
$K^0-\bar{K}^0$ mixing, which were discussed in Ref.~\cite{Rparity1}
at the leading-order, and the constraints on the relevant
combinations of couplings through comparing with data were given
by~\cite{Rparity1}
\begin{align}
  \sum_i \lambda'_{i31} \lambda'^*_{i13} n_{i} \lesssim
  3.3\times10^{-8}, \nonumber
  \\
  \sum_i \lambda'_{i21} \lambda'^*_{i12} n_{i} \lesssim
  4.5\times10^{-9},  \label{eq:old}
\end{align}
where $n_i \equiv (100~\text{GeV}/{m_{\tilde{\nu}_{iL}}})^2$,
$m_{\tilde{\nu}_{iL}}$ is the mass of the sneutrino of the $i$-th
generation. However, authors of Ref.~\cite{Rparity1} did not include
the contributions of the SM, since they believed that both
theoretical and experimental results involved considerable
uncertainties at that time. Recently, both SM theoretical
predictions and experimental results has been improved significantly
\cite{B0andK}. Thus, besides investigating of
$B^0_{s}$-${\bar{B}^0_{s}}$ mixing, it is also worthwhile to
reinvestigate the constraints on the combinations of R-parity
violating (RPV) couplings with the updated data on
$B^0_{d}$-${\bar{B}^0_{d}}$ and $K^0$-$\bar{K}^0$ mixing including
the SM contributions. Moreover, in general, the
next-to-leading-order (NLO) QCD corrections are significant, so we
also calculate the NLO QCD effects on the above neutral meson mixing
in this paper.

In addition to the above mass differences, the R-parity violating
couplings
  can also contribute to the CP asymmetries in the meson decay processes, when
  the couplings are complex. So the combinations of RPV couplings discussed above can affect
  observables related to time-dependent CP violation in processes such as $B_d \to J/\psi
  K_s$ and $B_s \to J/\psi \phi$. In this paper, we also consider the constraints on the
  relevant combinations of MSSM-RPV couplings from these CP violation observables.

We organize our paper as following. Section~\ref{sec:basic} is a
brief summary of the formalism for the calculation of the neutral
meson mixing , CP asymmetry in B physics and the results in the SM.
In Section~\ref{sec:nlo} , we calculate the contributions from the
MSSM-RPV at leading-order and next-to-leading-order in QCD. In
section~\ref{sec:num}, we present our numerical results and
discussions.

\section{Basic formalism and analytical results in the SM}
\label{sec:basic}

In order to make our paper self-contained, we first illustrate the
basic formalism for the calculation of the
$B^0_{s(d)}$-$\bar{B}^0_{s(d)}$ and $K^0$-$\bar{K}^0$ mass
differences $\Delta m$, the time-dependent CP asymmetry in $B_d$
decay, $S_{J/\psi K_s}$ and the time-dependent CP asymmetry in $B_s$
decay, $S_{J/\psi \phi}$.

We start from the $\Delta{F}=2$ effective Hamiltonian~\cite{buras1}
\begin{align}
  \mathcal{H}_{\text{eff}} = \sum_i C_i Q_i + h.c. .
  \label{effeham}
\end{align}
The relevant operators for our concerning are
\begin{align}
  Q_0 &= \bar{q}^\alpha \gamma_\mu P_L b^\alpha \bar{q}^\beta \gamma_\mu P_L b^\beta,
  \\
  Q_1 &= \bar{q}^\alpha P_L b^\alpha \bar{q}^\beta P_R b^\beta,
  \\
  Q_2 &= \bar{q}^\alpha P_L b^\beta \bar{q}^\beta P_R b^\alpha,
\end{align}
where $q=d,s$ for $B_d$ or $B_s$ meson mixing, respectively. Similar
expressions for $K^0-\bar{K}^0$ mixing can be obtained by replacing
$b$ with $s$ and setting $q=d$.

With this effective Hamiltonian, the $B_q-\bar{B}_q$ mass difference
can be expressed as
\begin{align}
  &\Delta m_q = 2 \left| \bra{\bar{B}_q} \mathcal{H}_{\text{eff}} \ket{B_q} \right| =
  2 \left| \sum_i C_i \bra{\bar{B}_q} Q_i \ket{B_q}
  \right|;
\end{align}
$a_{J/\psi K_s ( \phi )}$, time-dependent CP asymmetry in $B_{d( s
)} \rightarrow J/\psi K_s ( \phi )$ decays, can be expressed as
\begin{align}
  a_{J/\psi K_s ( \phi ) } = S_{J/\psi K_s ( \phi ) } \text{sin} \Delta m_{d(s)} t,
\end{align}
where $S_{J/\psi K_s ( \phi ) } = \text{sin} 2 \beta_{eff}$, and
$\beta_{eff} =\frac{1}{2} \text{arg} \bra{\bar{B}_{d(s)}}
\mathcal{H}_{\text{eff}} \ket{B_{d(s)}} $. In the framework of SM,
\begin{align}
  S^{SM}_{J/\psi K_s} = \text{sin} 2 \beta ,
  ~~S^{SM}_{J/\psi \phi} = \text{sin} 2 \beta_s ,
\end{align}
with
\begin{align}
  \beta = \text{arg}( -
  \frac{(V_{CKM})_{cd}(V_{CKM}^{\ast})_{cb}}{(V_{CKM})_{td}(V_{CKM}^{\ast})_{tb}}),
  \nonumber
  \\
  \beta_s = \text{arg}( -
  \frac{(V_{CKM})_{ts}(V_{CKM}^{\ast})_{tb}}{(V_{CKM})_{cs}(V_{CKM}^{\ast})_{cb}}).
\end{align}

In the SM, only $C_0$ is nonzero, which has been calculated to the
NLO in QCD~\cite{NLOQCDSM,buras1} and is given by
\begin{align}
  C_0(\mu) = \frac{G_F^2m_W^2}{4\pi^2} (V^*_{tb} V_{tq})^2 \eta_B S_0(x_t) \left(
    \alpha_s^{(5)}(\mu) \right)^{-6/23} \left[ 1 + \frac{\alpha_s^{(5)}(\mu)}{4\pi} J_5
  \right],
\end{align}
where $\eta_B=0.55\pm0.1$~\cite{etaparameter}, $J_5=1.627$ in the naive dimensional
regularization scheme (NDR), $x_t=m_t^2/m_W^2$ and $S_0(x_t)$ is the Inami-Lim
function~\cite{Inami-Lim}. The scale $\mu$ is usually taken to be $\sim m_b$ in $B$
physics.

The expression for the $K^0-\bar{K}^0$ mass difference is a little
different, which is given by
\begin{align}
  \Delta m_K = 2 \mathrm{Re} \bra{\bar{K}^0} \mathcal{H}_{\text{eff}} \ket{K^0} = 2
  \mathrm{Re} \sum_i C_i \bra{\bar{K}^0} Q_i \ket{K^0}.
  \label{K}
\end{align}
The nonzero NLO Wilson coefficient in the SM is
\begin{align}
  C_0(\mu) = \frac{G_F^2m_W^2}{4\pi^2} \left[ \lambda_c^2 \eta_1 S_0(x_c) + \lambda_t^2
    \eta_2 S_0(x_t) + 2 \lambda_c \lambda_t \eta_3 S_0(x_c,x_t) \right] \left(
    \alpha_s^{(3)}(\mu) \right)^{-2/9} \left[ 1 + \frac{\alpha_s^{(3)}(\mu)}{4\pi} J_3
  \right],
\end{align}
where $\lambda_i = V^*_{is} V_{id}$, $J_3=1.895$ in the NDR scheme
and the parameters $\eta_i$ are $\eta_1=1.38\pm0.20$,
$\eta_2=0.57\pm0.01$,
$\eta_3=0.47\pm0.04$~\cite{etaofK,etaparameter}, respectively. $S_0$
functions can be found in Ref.\cite{buras1}.

The matrix elements of the operators $Q_i$ (i=0,1,2) between two
hadronic states can be obtained by using the vacuum insertion
approximation (VIA) and partial axial current conservation (PCAC).
We refer the readers to Ref.~\cite{CP-violation} for details. The
results are
\begin{align}
  \braket{Q_0(\mu)} &\equiv \bra{\bar{B}_s} Q_0(\mu) \ket{B_s} = \frac{1}{3} f_{B_s}^2
  m_{B_s} B_0(\mu),
  \\
  \braket{Q_1(\mu)} &\equiv \bra{\bar{B}_s} Q_1(\mu) \ket{B_s} = f_{B_s}^2 m_{B_s} \left(
    \frac{1}{24} - \frac{m_{B_s}^2}{4(m_s(\mu)+m_b(\mu))^2} \right) B_1(\mu),
  \\
  \braket{Q_2(\mu)} &\equiv \bra{\bar{B}_s} Q_2(\mu) \ket{B_s} = f_{B_s}^2 m_{B_s} \left(
    \frac{1}{8} -\frac{m_{B_s}^2}{12(m_s(\mu)+m_b(\mu))^2} \right) B_2(\mu),
\end{align}
where $m_{B_s}$ and $f_{B_s}$ is the mass and decay constant of the
$B_s$ mesons, respectively. Clearly, the expressions for $B_d$ and
$K^0$ mesons can be obtained by simple substitutions. $B_i(\mu)$ is
the non-perturbative parameters and can be calculated with lattice
method~\cite{KBparameter,latticeB}. We list their values to be used
in our numerical calculations in table \ref{Bpara}:
\begin{table}
  \begin{tabular}{|c|c|c|c|}
    \hline & $B_0(\mu)$ & $B_1(\mu)$ & $B_2(\mu)$ \\
    \hline $K^0 (\mu=2 GeV)$ & 0.69 & 1.03 & 0.73 \\
    \hline $B_d (\mu=m_b)$ & 0.87 & 1.16 & 1.91 \\
    \hline $B_s (\mu=m_b)$ & 0.86 & 1.17 & 1.94 \\
    \hline
  \end{tabular}
  \caption{Non-perturbative parameters used in our numerical calculations.}
  \label{Bpara}
\end{table}

\section{MSSM-RPV contributions and NLO QCD corrections}
\label{sec:nlo}

In the MSSM-RPV, there are tree-level contributions to the
$B_{s(d)}-\bar{B}_{s(d)}$ and $K^0-\bar{K}^0$ mixing through
sneutrino exchange. The tree diagrams contribute through the
operator $Q_1$, and the corresponding Wilson coefficient is
\begin{align}
  C_1 = - \sum_i \frac{\lambda'_{ijk}\lambda'^*_{ikj}}{m_{\tilde{\nu}_i}^2} +
  \mathcal{O}(\alpha_s),
\end{align}
where $jk=32,31,21$ for $B_s$, $B_d$ and $K^0$ mesons, respectively.
We assume universal sneutrino masses for simplicity, i.e.,
$m_{\tilde{\nu}_i}=m_{\tilde{\nu}}$, $i=1,2,3$.

At NLO in QCD, both $Q_1$ and $Q_2$ contribute due to color
exchange. To calculate the NLO Wilson coefficients, we match the
full theory onto the effective theory at the SUSY scale, and then
run the coefficients down to the hadronic scales using the
renormalization group equations (RGE). In our calculations, we use
dimensional regularization in $d=4-2\epsilon$ dimensions to
regulate, and use $\overline{\text{MS}}$ scheme to renormalize the
ultraviolet (UV) divergences. We keep the heaviest quark mass $m_Q$
($Q=b$ for $B$ mesons and $Q=s$ for $K$ mesons) and set all other
quark masses to be zero in the Wilson coefficients. However, we will
keep the lighter quark mass $m_q$ in the intermediate stages of our
calculation in order to regulate the infrared (IR) divergences. We
also set all external momenta to zero, since the coefficients should
not depend on them.

\begin{figure}[ht!]
\includegraphics[scale=0.6]{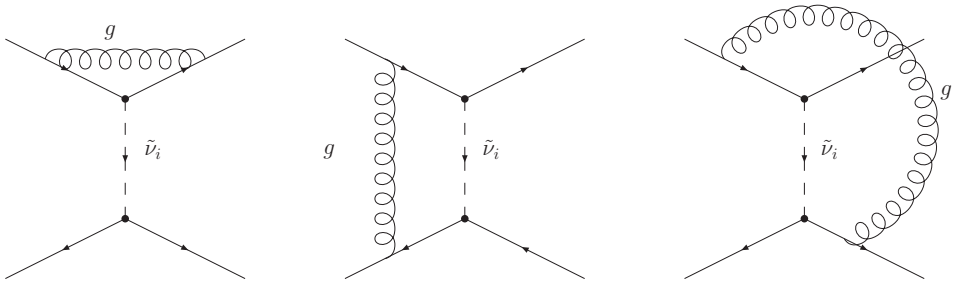}
\caption{One-loop Feynman diagrams in the full theory.}
\label{full}
\end{figure}

The NLO diagrams in the full theory are shown in Fig.~\ref{full} In the full theory, all
the UV divergences should be removed by the renormalization of the quark wave functions
and the coupling constants. The renormalized amplitude in the full theory is
\begin{align}
  A_{\text{full}} &= - \sum_i \frac{\lambda'_{ijk}\lambda'^*_{ikj}}{m^2_{\tilde{\nu}}}
  \left\{ \left[ 1 + \frac{\alpha_s}{\pi} C_F \left( 1 - 2\ln\frac{m_Q^2}{\mu^2} \right)
      + \frac{\alpha_s}{4\pi} \frac{ m_{\tilde{\nu}}^2 \ln\frac{m_Q^2}{m_q^2} + m_Q^2
        \ln\frac{m_q^2}{m_{\tilde{\nu}}^2} } {m_{\tilde{\nu}}^2-m_Q^2} \right]
    \braket{Q_1}_{\text{tree}} \right. \nonumber
  \\
  &\qquad \left. - \frac{\alpha_s}{4\pi} \frac{1}{N} \frac{ m_{\tilde{\nu}}^2
      \ln\frac{m_Q^2}{m_q^2} + m_Q^2 \ln\frac{m_q^2}{m_{\tilde{\nu}}^2} }
    {m_{\tilde{\nu}}^2-m_Q^2} \braket{Q_2}_{\text{tree}} \right\},
\end{align}
where $C_F=4/3$, $N=3$ is the number of colors, and $\braket{Q_i}_{\text{tree}}$ is the
tree level matrix elements of the operators.

\begin{figure}[ht!]
\includegraphics[scale=0.6]{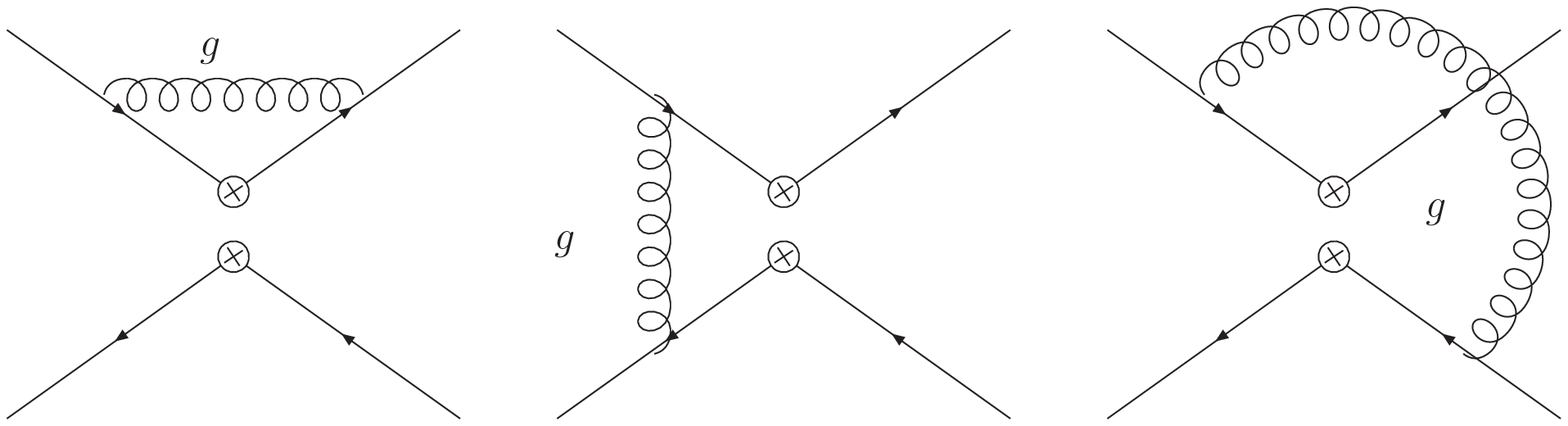}
\caption{The next-to-leading-order corrections in the effective theory.}
\label{effective}
\end{figure}

In the effective theory, there are remaining UV divergences after
taking into account the quark field renormalization, which must be
canceled by the renormalization of the effective operators.
Calculating the diagrams in Fig.~\ref{effective} with the insertion
of the operators $Q_1$ and $Q_2$, we get the following amplitudes
after quark field renormaliztion:
\begin{align}
  \braket{Q_1^0} &= \left[ 1 + \frac{\alpha_s}{2\pi} C_F \left( \frac{3}{\hat{\epsilon}}
      + 2 - 4\ln\frac{m_Q^2}{\mu^2} \right) + \frac{\alpha_s}{4\pi}
    \ln\frac{m_Q^2}{m_q^2} \right] \braket{Q_1}_{\text{tree}} - \frac{\alpha_s}{4\pi}
  \frac{1}{N} \ln\frac{m_Q^2}{m_q^2} \braket{Q_2}_{\text{tree}},
  \\
  \braket{Q_2^0} &= \frac{\alpha_s}{4\pi} \left( \frac{3}{\hat{\epsilon}} + 1 -
    3\ln\frac{m_Q^2}{\mu^2} - \frac{1}{N} \ln\frac{m_Q^2}{m_q^2} \right)
  \braket{Q_1}_{\text{tree}} \nonumber
  \\
  &\qquad + \left[ 1 - \frac{\alpha_s}{4\pi} \frac{1}{N} \left( \frac{3}{\hat{\epsilon}}
      + 1 - 3\ln\frac{m_Q^2}{\mu^2} \right) + \frac{\alpha_s}{2\pi} C_F \left( 1 -
      \ln\frac{m_Q^2}{\mu^2} \right) + \frac{\alpha_s}{4\pi} \ln\frac{m_Q^2}{m_q^2}
  \right] \braket{Q_2}_{\text{tree}},
\end{align}
where $1/\hat{\epsilon}=1/\epsilon-\gamma_E+\ln{4\pi}$. The renormalization constant
matrix for the two operators is
\begin{align}
  \hat{Z}_Q = 1 + \frac{\alpha_s}{4\pi} \frac{1}{\hat{\epsilon}}
  \begin{pmatrix} 8 & 0 \\ 3 & -1 \end{pmatrix},
\end{align}
from which we obtain the anomalous dimension matrix:
\begin{align}
  \hat{\gamma}_Q = - \frac{\alpha_s}{2\pi}
  \begin{pmatrix} 8 & 0 \\ 3 & -1 \end{pmatrix}.
\end{align}
Our results of anomalous dimensions agree with those in Ref.~\cite{general anomalous
  dimension}.

Matching the results in the full theory and the effective theory, we extract the Wilson
coefficients:
\begin{align}
  C_1(m_{\tilde{\nu}}) &= - \sum_i
  \frac{\lambda'_{ijk}\lambda'^*_{ikj}}{m^2_{\tilde{\nu}}} \left( 1
  +
    \frac{\alpha_s}{4\pi} \frac{x\ln {x}}{1-x} \right),
  \\
  C_2(m_{\tilde{\nu}}) &= ~~ \sum_i
  \frac{\lambda'_{ijk}\lambda'^*_{ikj}}{m^2_{\tilde{\nu}}} \frac{\alpha_s}{4\pi}
  \frac{1}{N} \frac{x\ln {x}}{1-x},
\end{align}
where $x=m_Q^2/m^2_{\tilde{\nu}}$. These coefficients satisfy the renormalization group
equations
\begin{align}
  \frac{d}{d\ln\mu} \begin{pmatrix} C_1(\mu) \\ C_2(\mu) \end{pmatrix} =
  \hat{\gamma}_Q^T(\alpha_s(\mu))
  \begin{pmatrix} C_1(\mu) \\ C_2(\mu) \end{pmatrix},
\end{align}
from which we can solve the Wilson coefficients for arbitrary scale $\mu$.

$\Delta m_{B}$, $S_{J/\psi K_s(\phi)}$ are defined in terms of the
matrix element $\bra{\bar{B}_{s(d)}} \mathcal{H}^{\Delta B =
2}_{eff} \ket{B_{s(d)}}$, which can be written as

\begin{equation}
\bra{\bar{B}_{s(d)}} \mathcal{H}^{\Delta B = 2}_{eff} \ket{B_{s(d)}}
= \mathcal{A}_{SM} + \mathcal{A}_{RPV} = \mathcal{A}_{SM} (1 +
\frac{\mathcal{A}_{RPV}}{\mathcal{A}_{SM}} ),
\end{equation}
where $\mathcal{A}_{SM}$ and $\mathcal{A}_{RPV}$ denote matrix
elements of the SM and the MSSM-RPV effective hamiltanian,
respectively. With the above matrix element, $\Delta m_{B}$ and
$S_{J/\psi K_s(\phi)}$ can be expressed
\begin{align}
&\Delta m_{B} = 2 \left| \bra{\bar{B}_{s(d)}} \mathcal{H}^{\Delta B
= 2}_{eff} \ket{B_{s(d)}} \right| = \Delta m^{SM}_{B} \left| 1 +
\frac{\mathcal{A}_{RPV}}{\mathcal{A}_{SM}} \right|, \nonumber
\\
&S_{J/\psi K_s(\phi)} = \text{sin} \left( 2 \beta_{(s)} + \text{arg}
\left( 1 + \frac{\mathcal{A}_{RPV}}{\mathcal{A}_{SM}}\right)
\right), \label{basicformula}
\end{align}
where $\Delta m^{SM}_{B}$ and $\beta_{(s)}$ denotes the SM
contributions, respectively. In Eq.(\ref{basicformula}), hadronic
uncertainties arising from hadron decay constants cancel between
$\mathcal{A}_{SM}$ and $\mathcal{A}_{RPV}$, and hadronic
uncertainties remain only in $\Delta m^{SM}_{B}$ and $\beta_{(s)}$.
As for $K^0-\bar{K}^0$ mixing, $\Delta m_K$ can be obtained
straightforwardly.

\section{Numerical results}
\label{sec:num}

In this section we present our numerical results. The SUSY scale is
taken to be the mass of the sneutrino $m_{\tilde{\nu}}$. The CKM
matrix elements are parametrized in the Wolfenstein convention with
four parameters $A=0.809$, $\lambda=0.2272$, $\rho=0.197$ and
$\eta=0.339$. The other standard model parameters are taken to be
$G_F=1.16637^{-5}$~GeV$^{-2}$, $\alpha_s(m_Z)=0.118$,
$m_t=173$~GeV~\cite{Bparameter}. The mass of B meson and K meson are
$m_{B_s}=5367.5$~MeV, $m_{B_0}=5279.4$~MeV and
$m_{K}=497.648$~MeV~\cite{Bparameter}. The time-dependent CP
asymmetry in $B_d$ decay $S_{J/\psi
K_s}=0.687^{+0.032}_{-0.032}$~\cite{Bparameter}. The recent
experimental values of the mass differences of the $B_d$ and $K^0$
mesons mixing are
$\Delta{m_K}=(0.5292\pm0.0009)\times10^{-2}$~ps$^{-1}$,
$\Delta{m_{B_d}}=0.507^{+0.005}_{-0.005}$~ps$^{-1}$~\cite{Bparameter},
and $\Delta{m_s}$ is shown in Eq.~(\ref{FBsmixing}).

In our numerical calculations, we first neglect the uncertainties
from the hadronic parameters in the SM and assume that the
predictions of the SM can reproduce the central values of the
experimental data. Thus we demand that the RPV contributions must
not exceed the experimental upper and lower bounds of the
corresponding data.

\begin{figure}
\includegraphics[scale=1.2]{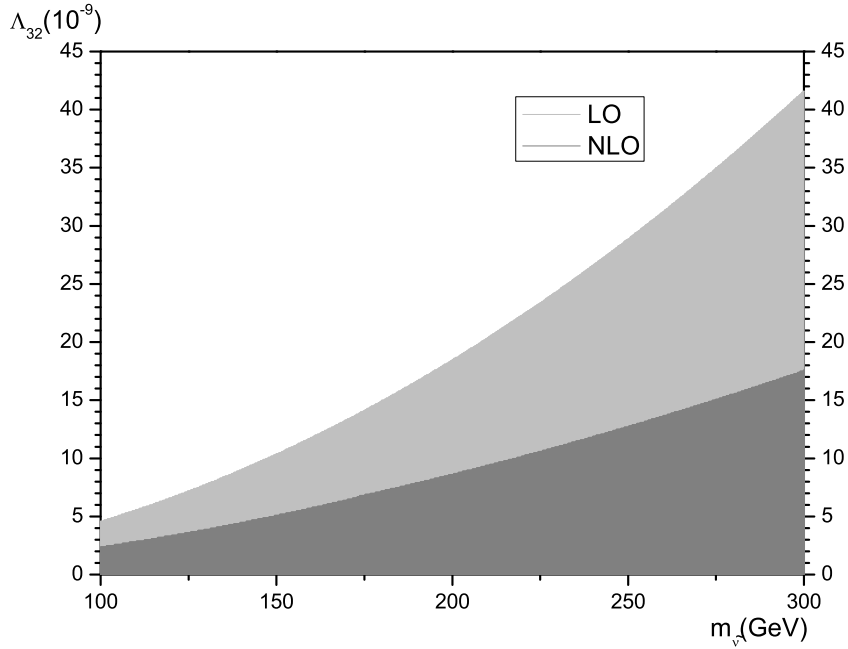}
\caption{\label{diagram1}The bounds for
$\sum\limits_i\lambda'_{i32}\lambda'^*_{i23}$
  vary with the mass of the sneutrino. The gray area and the dark area are the allow
  region extracted from the LO amplitude. After NLO corrections, the gray area is
  excluded and only the dark area survives.}
\end{figure}

First, we consider the new measurement of $B_s-\bar{B}_s$ mixing,
which can constrain the combination
$\Lambda_{32}\equiv\sum\limits_i\lambda'_{i32}\lambda'^*_{i23}$ once
the sneutrino mass is given. Fig.~\ref{diagram1} shows the allowed
region in the $\Lambda_{32}-m_{\tilde{\nu}}$ plane, where the gray
area and the dark area is the allowed region extracted from the
leading-order amplitude. We find that with the new data, the
couplings are constrained to the level of $10^{-9}$. After including
the NLO QCD corrections, the gray area is excluded and only the dark
area survives. One can see that since the NLO QCD effects increase
the resulting mass difference, the bound is more stringent (about
50\% lower) than that from LO calculations. If setting
$m_{\tilde{\nu}}=100$~GeV, the bounds on $\Lambda_{32}$ from the
calculations of LO and NLO are
\begin{align}
 &\Lambda_{32} < 4.6\times10^{-9}\text{(LO)}, \quad
 2.4\times10^{-9}\text{(NLO)} ,
 \label{constraintsBs}
\end{align}
respectively.

\begin{figure}
\includegraphics[scale=1.2]{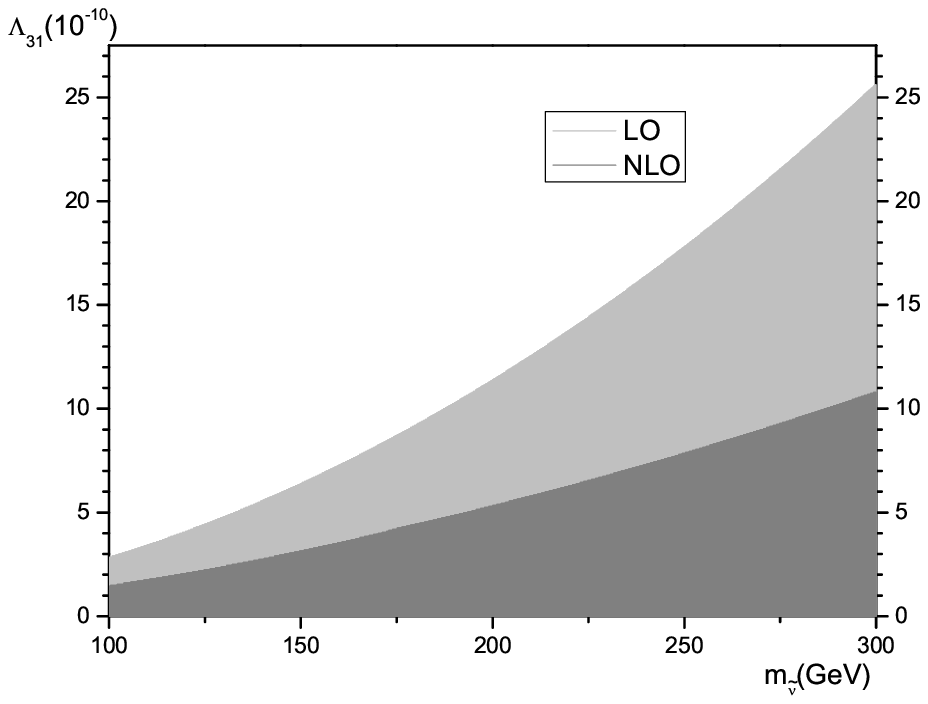}
\caption{\label{diagram2}The bounds for
$\sum\limits_i\lambda'_{i31}\lambda'^*_{i13}$
  vary with the mass of the sneutrino.}
\end{figure}

\begin{figure}
\includegraphics[scale=1.2]{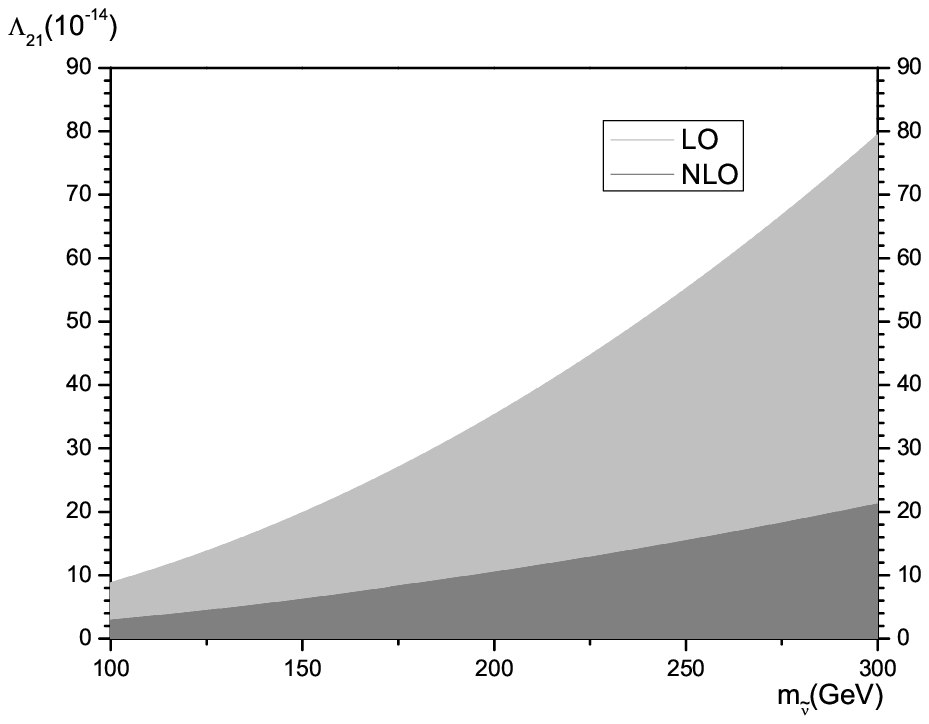}
\caption{\label{diagram3}The bounds for
$\sum\limits_i\lambda'_{i21}\lambda'^*_{i12}$
  vary with the mass of the sneutrino.}
\end{figure}

Fig.~\ref{diagram2} and Fig.~\ref{diagram3} show the constraints on
the combination
$\Lambda_{31}\equiv\sum\limits_i\lambda'_{i31}\lambda'^*_{i13}$ from
$B_0-\bar{B}_0$ mixing and
$\Lambda_{21}\equiv\sum\limits_i\lambda'_{i21}\lambda'^*_{i12}$ from
$K_0-\bar{K}_0$ mixing. The situation here is similar to that in
$B_s-\bar{B}_s$ mixing: the NLO QCD corrections increase the mass
differences and thus give more strong constraints on the
combinations of couplings of MSSM-RPV. At the reference point
$m_{\tilde{\nu}}=100$~GeV, the bounds are
\begin{align}
  &\Lambda_{31} < 2.9\times10^{-10}\text{(LO)}, \quad
  1.5\times10^{-10}\text{(NLO)},
  \label{constraintsBd}
  \\
  &\Lambda_{21} < 8.8\times10^{-14}\text{(LO)}, \quad 2.8\times10^{-14}\text{(NLO)}.
  \label{constraintsK}
\end{align}
Above bounds are two and four orders of magnitude stronger than ones
given in Ref.~\cite{Rparity1}, as shown in Eq.~(\ref{eq:old}), which
is due to the fact that the updated data of the measurements and the
contributions from the SM have been considered in our calculations.

We further consider the situation that the hadronic
  uncertainties are involved. In this case, the bounds on the combinations of couplings
  in MSSM-RPV are looser than those in Eqs. (\ref{constraintsBs}), (\ref{constraintsBd})
  and (\ref{constraintsK}). The new bounds from the three observables are
\begin{align}
  \Lambda_{32} &< 12.392\times10^{-8}  \quad\text{(LO)},
  \nonumber
  \\
  &< 6.481\times10^{-8} \quad\text{(NLO)} ;
  \nonumber
  \\
  \Lambda_{31} &< 7.833\times10^{-9}  \quad\text{(LO)},
  \nonumber
  \\
  &< 4.096\times10^{-9}\quad\text{(NLO)};
  \nonumber
  \\
  \Lambda_{21} &< 3.47\times10^{-11}   \quad\text{(LO)},
  \nonumber
  \\
  &< 1.39\times10^{-11} \quad\text{(NLO)} ,
\end{align}
respectively.

\begin{figure}
\includegraphics[scale=1.2]{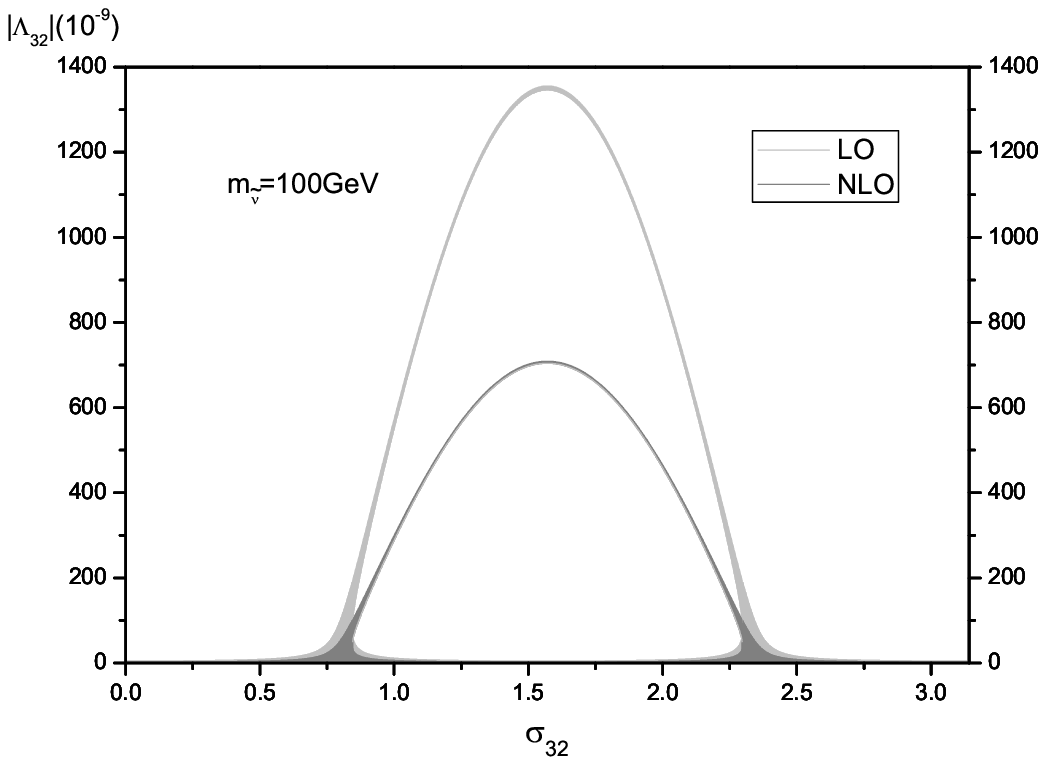}
\caption{\label{diagramBsc}The allowed range of the magnitude and
phase of $\sum\limits_i\lambda'_{i32}\lambda'^*_{i23}$ constrained
by the updated experimental data of $\Delta m_s$. The gray area
accounts for the tree level results. When the QCD corrections are
added, the allowed area is reduced to the dark area.  }
\end{figure}

\begin{figure}
\includegraphics[scale=1.2]{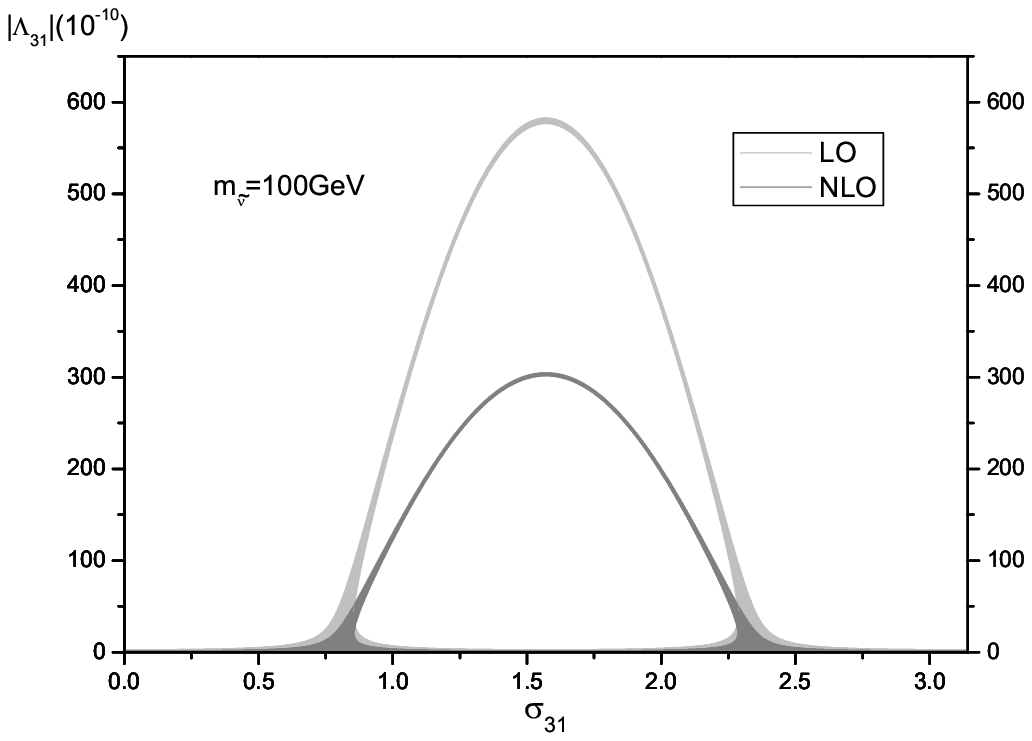}
\caption{\label{diagramBsd}The allowed range of the magnitude and
phase of $\sum\limits_i\lambda'_{i31}\lambda'^*_{i13}$ constrained
by the experimental data of $\Delta m_d$.}
\end{figure}

We also discuss the situation that $\Lambda_{ij}$ is complex, and
parametrize the combinations of relevant couplings as $\Lambda_{ij}
= |\Lambda_{ij}| e ^{2i\sigma_{ij}}$, where $ij$ can be $32$ and
$31$. After assuming arbitrary phases $\sigma_{ij}$, we can obtain
constraints on the magnitude $|\Lambda_{ij}|$. Fig.~\ref{diagramBsc}
and Fig.~\ref{diagramBsd} shows the the allowed range of
$\Lambda_{32}$ and $\Lambda_{31}$ for $m_{\tilde{\nu}} =  100$GeV,
taking into account the constraints from $\Delta m_s$ and $\Delta
m_d$. For the $B_s-\bar{B}_s$ mixing, $\Lambda_{32}$ is about
$\mathcal{O}(10^{-9})$, while for the $B_d-\bar{B}_d$ mixing,
$\Lambda_{31}$ is about $\mathcal{O}(10^{-11})$. In both figures,
there are additional areas with large $|\Lambda_{ij}|$ besides
ordinary areas with small $|\Lambda_{ij}|$, which correspond to ones
where contributions from the MSSM-RPV are larger than those from the
SM, roughly twice the SM contributions but with different signs. We
do not discuss complex couplings in K system due to the fact that
Eq. (\ref{K}) holds only under the condition that imaginary part of
$\bra{\bar{K}^0} \mathcal{H}_{\text{eff}} \ket{K^0}$ is far less
than the real part.

\begin{figure}
\includegraphics[scale=1.2]{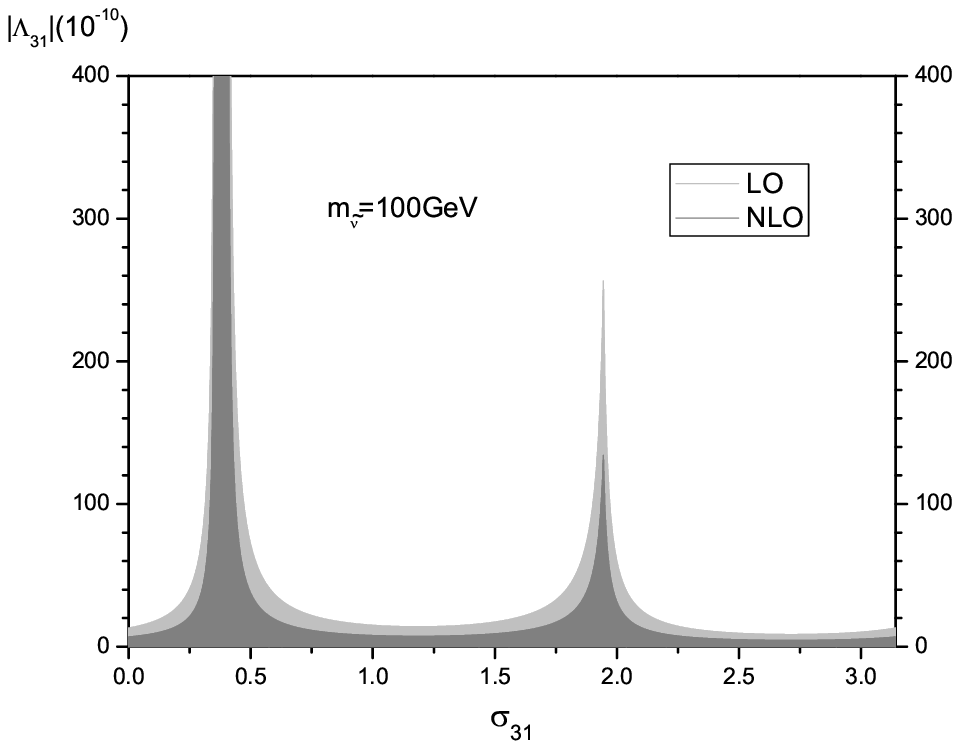}
\caption{\label{sinbeta}The allowed range of the magnitude and phase
of $\sum\limits_i\lambda'_{i31}\lambda'^*_{i13}$ using experimental
data of $S_{J/\psi K_s}$.}
\end{figure}

\begin{figure}
\includegraphics[scale=1.2]{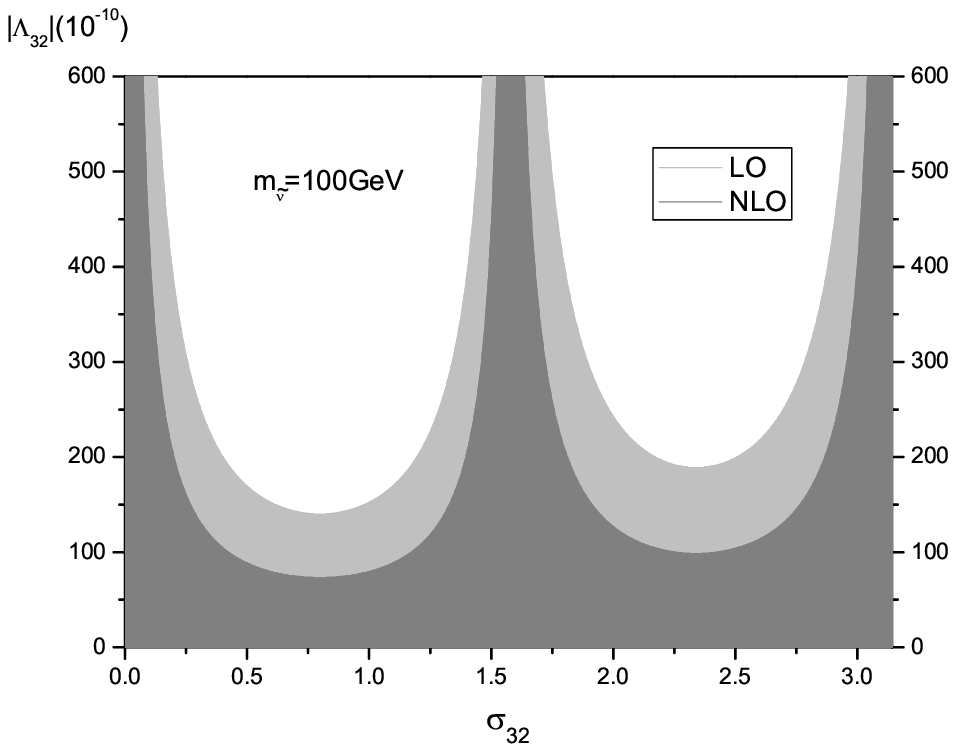}
\caption{\label{sinbetas}The allowed range of the magnitude and
phase of $\sum\limits_i\lambda'_{i32}\lambda'^*_{i23}$, assuming the
first year LHCb data provide an experiment measurement $S_{J/\psi
\phi} = 0.04 \pm 0.03$.}
\end{figure}

The combinations of complex couplings can introduce new CP-violation
  origins, so we further investigate how the CP asymmetries in B decays constrain these
  combinations of couplings. Using the experimental data of $S_{J/\psi
    K_s}$~\cite{Bparameter}, we plot the allowed areas of the magnitude $|\Lambda_{31}|$
  and the phase $\sigma_{31}$ in Fig.\ref{sinbeta}, which shows that the upper bound for
  $|\Lambda_{31}|$ is generally about $\mathcal{O}(10^{-10})$, except for some special values of
  $\sigma_{31}$. Those special values correspond to
$\arg(1+\mathcal{A}_{RPV}/\mathcal{A}_{SM}) = 0$, and we have
  $\sin(2\beta_{\text{eff}})=\sin(2\beta)$.
In fact, another area with
$\arg(1+\mathcal{A}_{RPV}/\mathcal{A}_{SM}) = \pi-4\beta$ also
survives from constraint of experiment data of $S_{J/\psi K_s}$,
however, this possibility has been excluded by an angular analysis
of $B^0 \rightarrow J/\psi K^{*0}$ and a time-dependent Dalitz plot
analysis of $B^0 \rightarrow \bar{D}^0h^0 (h^0 = \pi^0, \eta^0,
\omega)$\cite{piminus2beta}.
  For the CP asymmetry in $B_s$ decays,
  although there is no experimental data of $S_{J/\psi \phi} $ currently, future LHCb
  experiment will provide enough data to reach $\sigma_{stat}(S_{J/\psi \phi})\approx
  0.03$ in the first year~\cite{LHCb}, which would provide a strong constraint on new
  physics. So we assume $S_{J/\psi \phi} = 0.04 \pm 0.03$ and plot constraints on
  $|\Lambda_{32}|$ and $\sigma_{32}$ in Fig.\ref{sinbetas}, which shows that the upper
  bound for $|\Lambda_{32}|$ is about $\mathcal{O}(10^{-8})$ except for some special
  values of $\sigma_{32}$ similar to the case of $S_{J/\psi K_s}$.

In conclusion, using the updated data, we have calculated
contributions to $B_s-\bar{B}_s$ mixing, $B_d-\bar{B}_d$ mixing and
$K^0-\bar{K}^0$ mixing in the framework of the minimal
supersymmetric standard model with R-parity violation including NLO
QCD corrections, and presented new constraints on the relevant
couplings of MSSM-RPV. Our results show that upper bound on the
relevant combination of couplings of $B_s-\bar{B}_s$ mixing is of
the order $10^{-9}$, and upper bounds on the relevant combinations
of couplings of $B_d-\bar{B}_d$ mixing and $K^0-\bar{K}^0$ mixing
are two and four orders of magnitude stronger than ones reported in
Ref.~\cite{Rparity1}, respectively. We also discussed the case of
complex couplings and showed that how the relevant combinations of
couplings are constrained by the updated experiment data of
$B_s-\bar{B}_s$, $B_d-\bar{B}_d$ mixing and time-dependent CP
asymmetry $S_{J/\psi K_s}$, and future possible observations of
$S_{J/\psi\phi}$ at LHCb, respectively.

$Note~added$. While preparing this manuscript the paper of
\cite{RPVnew} appeared where the same coupling combination from
$B_s-\bar{B}_s$ mixing is also discussed. However, the authors
of~\cite{RPVnew} mainly dealt with contributions through the box
diagram. Our results induced by the tree-level diagram is a few
orders of magnitude stronger than theirs.

\begin{acknowledgments}
This work was supported in part by the National Natural Science
Foundation of China, under Grant No.~10421503, No.~10575001 and
No.~10635030, and the Key Grant Project of Chinese Ministry of
Education under Grant No.~305001 and the Specialized Research Fund
for the Doctoral Program of Higher Education.
\end{acknowledgments}

\end{document}